\documentclass[12pt]{article}

\usepackage[utf8]{inputenc}
\usepackage{amsmath,amssymb}
\usepackage{slashed}
\usepackage{authblk}
\usepackage{xcolor}
\usepackage{hyperref}

\usepackage[dvips]{graphicx}
\usepackage{cite}

\newcommand{\sss}[1]{\scriptscriptstyle{#1}}
\newcommand{\sqrtL}{\sqrt{\lambda_e}}
\newcommand{\SANC}{\texttt{SANC}}
\newcommand{\MCSANC}{\texttt{MCSANC}}
\newcommand{\ReneSANCe}{\texttt{ReneSANCe}}
\newcommand{\WHIZARD}{\texttt{WHIZARD}}
\newcommand{\CalcHEP}{\texttt{CalcHEP}}

\newcommand{\mw}{M_{\sss W}} 
\newcommand{\mz}{M_{\sss Z}}
\newcommand{\ml}{m_l}
\newcommand{\stw}{s_{\sss W}}
\newcommand{\ctw}{c_{\sss W}}
\newcommand{\tffgg} {{\tilde F}_{\sss gg}}
\newcommand{\tffld}{{\tilde F}_{\sss ld}}
\newcommand{\tffqd}{{\tilde F}_{\sss qd}}
\newcommand{\tffqq}{{\tilde F}_{\sss qq}}
\newcommand{\tffll}{{\tilde F}_{\sss ll}}
\newcommand{\tfflq}{{\tilde F}_{\sss lq}}

\newcommand{\mzs}{M^2_{\sss{Z}}}

\newcommand{\qel}{Q_e}

\newcommand{\cpl}{c^+}
\newcommand{\cml}{c^-}

\newcommand{\vmalep}{\delta_{l}}

\newcommand{\alep}{I^{(3)}_l}

\newcommand{\chizt}{\chi_{\sss{Z}}(t)}
\newcommand{\nll}{\nonumber\\}
\newcommand{\Litwo}{{\rm{Li}}_{2}}
\newcommand{\bqa}{\begin{eqnarray}}
\newcommand{\eqa}{\end{eqnarray}}
\newcommand{\ds }{\displaystyle}

\newcommand{\rlms}{m_l^2}
\newcommand{\lnlaom}{\ln\left(\frac{ 4 \omega^2}{\lambda^2}\right)}

\newcommand{\sqs }{\sqrt{s}}

\usepackage{scalerel}

\def\nll{\nonumber\\}
\def\ds{\displaystyle}

\begin{document}

\title{
One-loop electroweak radiative corrections to polarized M{\o}ller 
scattering}

\author[1]{S.G.~Bondarenko}
\author[2]{L.V.~Kalinovskaya}
\author[2]{L.A.~Rumyantsev}
\author[2,3]{V.L.~Yermolchyk}

\affil[1]{\small Bogoliubov Laboratory of Theoretical Physics, Joint Institute for Nuclear Research, Dubna, 141980 Russia}
\affil[2]{\small Dzhelepov Laboratory of Nuclear Problems, Joint Institute for Nuclear Research, Dubna, 141980 Russia}
\affil[3]{\small Institute for Nuclear Problems, Belarusian State University,
Minsk, 220006  Belarus}

\maketitle

\begin{abstract}
This work is devoted to a theoretical description of polarized M{\o}ller scattering.
Complete one-loop electroweak radiative corrections are calculated 
in the helicity amplitude approach with allowance for the exact dependence on the muon mass.
Numerical results are presented for integrated unpolarized and polarized cross sections
as well as angular differential distributions.
Calculations are performed using
\ReneSANCe{} Monte Carlo ~generator
and  \MCSANC{} Monte Carlo integrator.
\end{abstract}

\section{Introduction}\label{Sect1}

The next generation of electron colliders -- 
the  International  Linear  Collider(ILC)~\cite{homepagesILC,Irles:2019xny,
  Moortgat-Picka:2015yla,Baer:2013cma,Accomando:1997wt,Battaglia:2004mw},
the $e^+e^-$ Future  Circular  Collider  (\texttt{FCCee})
\cite{homepagesFCCee,Abada:2019ono,Abada:2019lih,Blondel:2019ykp,Blondel:2018mad},
the  Compact  Linear  Collider  (\texttt{CLIC}) \cite{homepagesCLIC,CLIC:2016zwp,Charles:2018vfv},
and the  Circular  Electron  Positron  Collider  (\texttt{CEPC}) \cite{homepagesCEPC} -- will allow  an extensive program of experiments with unique opportunities for precision measurements. A major advantage to fulfill this goal is the universality of linear colliders, as they can operate in four $e^+e^-$, $e^-e^-$, $e^-\gamma$
and $\gamma\gamma$ modes with strongly polarized electron and photon beams. 
An important feature of linear colliders is a high degree of polarization which can be obtained for electron beams. 

The M{\o}ller scattering along with the Bhabha and Compton-like processes
is a good candidate for luminosity measurements and the background
estimation
in many searches for new physics
beyond the Standard Model.
At high energies for the polarized M{\o}ller scattering the most advanced Monte Carlo tool is needed not only to estimate luminosity, i.e. polarized experiments \texttt{CLIC}~\cite{CLIC:2016zwp}, \texttt{ILC}~\cite{Fujii:2020pxe}, but also to study muon-muon polarized scattering at $\mu$\texttt{TRISTAN} \cite{Hamada:2022mua}.

Equal lepton scattering, $e^- e^- \to e^- e^-$ was first calculated by  C.~M{\o}ller in 1932 ~\cite{Moller1932}.
There are a great number of theoretical works for description
polarized case of this process 
\cite{Jadach:1994im,
Shumeiko:1999zd,
Montero:1998ve,
Denner:1998um,
Czarnecki:2000ic,
Alexander:2000bu,
Petriello:2002wk,
Ilyichev:2005rx}.
In this series of papers the calculations are given
for the QED and electroweak (EW) one-loop corrections with taking into account the polarization.

A calculation of the radiative corrections (RCs) was performed for the unpolarized M{\o}ller scattering for the experiment~\cite{MOLLER:2014iki}
at one-loop level \cite{Aleksejevs:2010ub,Ahmadov:2012se},
partly at two-loop level \cite{Aleksejevs:2015zya}
and in the first time
beyond the ultra-relativistic approximation in \cite{Akushevich:2015toa}.

However, all of the above-mentioned studies
are not accompanied by the development of the Monte Carlo event generator
which is the standard of the
modern theoretical support of the high-precision experiments.

The following Monte Carlo generators currently exist, which take into account polarization at tree level:
{\tt AMEGIC++} ~\cite{Krauss:2001iv}, based on the helicity amplitudes and being a part of {\tt SHERPA};
{\tt COMPHEP}~\cite{Belyaev:2012qa}, using the traditional trace techniques to evaluate the matrix elements;
{\tt GRACE}~\cite{Yuasa:1999rg,Belanger:2003sd} (with the packages {\tt BASES} and {\tt SPRING}), calculating matrix elements via helicity amplitude techniques;
{\tt WHIZARD} ~\cite{Kilian:2007gr} a software system, intended for the effective calculation of scattering
cross-sections of many-particle and simulated events, where polarization is processed for both the
initial and final states.

Theoretical support of experiments by the {\tt MERADGEN} MC generator for polarized M{\o}ller scattering within QED theory is presented in ~\cite{Afanasev:2006xs}.

In our previous works we estimated the theoretical uncertainty
for the complete one-loop and leading higher-order EW corrections
for $e^+e^-$ and $\gamma\gamma$ polarized beams. 
The implementations of polarized
Bhabha scattering \cite{Bardin:2017mdd}, 
polarized $e^+e^- \to ZH$
\cite{Bondarenko:2018sgg},
$s$ -channel 
\cite{Bondarenko:2020hhn},
 $e^+e^- \to \gamma Z$
\cite{Bondarenko:2021eni} and 
$\gamma\gamma \to ZZ$ \cite{Bondarenko:2022ddm}
are available in the \ReneSANCe{} MC generator ~\cite{Sadykov:2020any} and  the \MCSANC{} integrator in the fully massive case and in total phase space.

This article is the next step in the series of \SANC{}~ papers devoted to the implementation of one of the channels $4f \to 0$, namely,
the equal lepton scattering at the one-loop level
with allowance for polarization. 

The $\alpha(0)$  EW scheme is used in the calculations.
All the results are obtained 
for the center-of-mass system (c.m.s.) energies from $\sqrt{s}=250$~GeV up to $3$~TeV.
The sensitivity to the initial polarization for the Born and hard photon bremsstrahlung cross sections was estimated for four beam polarization data sets:
\bqa
&&(P_{e^-}, P_{e^{-}}) =\\
&&(0,0),(-1,-1),(-1,+1),(+1,-1),(+1,+1).
\nonumber
\eqa
The one-loop contributions were calculated for the following
degrees of polarization:
\bqa
(P_{e^-}, P_{e^{-}}) =
\label{SetPolarization2} 
(0,0),(\pm 0.8, \pm 0.8).
\eqa

The statistical uncertainties were estimated using the \SANC{} tools: \ReneSANCe{}~
MC generator  and ~\MCSANC{} integrator.

This article consists of four Sections.

We describe the methodology of calculations of the polarized cross sections 
at the complete one-loop EW level 
within the helicity approach in Section~\ref{Sect2}
Numerical
results and comparison are presented in the next Section~\ref{Sect_Num}. 
~Summary is drawn in Section~\ref{Sect_C}.

\section{EW one-loop radiative corrections}
\label{Sect2}

We consider the differential cross section
for  processes 
\bqa
\label{lepton-pm}
l^{\pm}(p_1,\chi_1)\ +\ l^{\pm}(p_2,\chi_2)\ \rightarrow \hskip 35mm \\
\hskip 20mm l^{\pm}(p_3,\chi_3)\ +\ l^{\pm}(p_4,\chi_4)\ \ (+ \gamma (p_5, \chi_5)),
\nonumber
\eqa
with $l=e,\mu$ and arbitrary longitudinal polarization of initial particles 
($\chi$ corresponds to the helicity of the particles).

Within the \SANC{} system we calculate all processes using the on-mass-shell renormalization
scheme in two gauges: the $R_\xi$ gauge
and the unitary gauge as a cross-check.

We apply the helicity approach (HA) to all components
of the one-loop cross sections:
\bqa
\sigma^{\text{one-loop}} = \sigma^{\mathrm{Born}} + \sigma^{\mathrm{virt}}(\lambda)
+ \sigma^{\mathrm{soft}}(\lambda,\omega) + \sigma^{\mathrm{hard}}(\omega),
\eqa
where $\sigma^{\mathrm{Born}}$ is the Born cross section,
$\sigma^{\mathrm{virt}}$ is the contribution of virtual (loop) corrections,
$\sigma^{\mathrm{soft(hard)}}$ is the soft (hard) photon emission contribution
(the hard photon energy {$E_{\gamma} > \omega$}).
The auxiliary parameters $\lambda$ ("photon mass")
and $\omega$ are canceled after summation.
The corresponding expressions for the M{\o}ller scattering cross section cannot be integrated
over all angles because the integral diverges at $\vartheta =0,\pi$.

\subsection{Born and virtual parts}

To calculate the virtual part
at the one-loop level using the procedure basement of \SANC{},
we start with considering the covariant amplitude (CA).
The covariant one-loop amplitude corresponds to the result of the
straightforward standard calculation  
of all diagrams contributing to a given process at the 
one-loop  level. 
The CA is represented in a certain basis made of strings of Dirac
matrices and/or  
external momenta (structures) contracted with polarization vectors of vector
bosons, $\epsilon(k)$, if any.  

CA can be written in an explicit form using scalar form factors (FFs). 
All masses, kinematical factors and coupling constant and other parameter
dependences are included into these FFs ${\cal F}_{i}$, but tensor structures
with Lorenz indices made of strings of Dirac matrices are given by the basis.

The number of FFs is equal to the number of independent structures.

Loop integrals are expressed in terms of standard scalar
Passarino-Veltman functions $ A_0, \, B_0, \, C_0, \, D_0 $~\cite{Passarino:1978jh}.
We presented the CA for the
 $4f \to 0$  process  in \cite{Andonov:2002xc},
where we considered it at the one-loop level of annihilation into a vacuum. 
Recall that in \SANC ~we always calculate any 
one-loop process amplitude as annihilation into vacuum with
all 4 momenta incoming. Therefore, the derived universal scalar form factors for the amplitude of the process $4f \to 0$ after
an appropriate permutation of their arguments can be
used for the description of the next-to-leading (NLO) corrections of 
this particular case  unfolding  
into $t$ and $u$ channels.

The virtual (Born) cross section of processes (\ref{lepton-pm})
can be written as follows:
\bqa
\frac{d\sigma^{\mathrm{virt(Born)}}_{ \chi_1 \chi_2}}{d\cos{\vartheta_{3}}}
= \pi\alpha^2\frac{\beta_s}{2s}|\mathcal{H}^{\mathrm{virt(Born)}}_{\chi_1 \chi_2}|^2,
\eqa
where
\bqa
|\mathcal{H}^{\rm virt(Born)}_{ \chi_1 \chi_2}|^2 =
\sum_{\chi_3,\chi_4} |\mathcal{H}^{\rm virt(Born)}_{\chi_1 \chi_2\chi_3\chi_4}|^2,
\eqa
where $m_l$ is the final lepton mass and $\beta_s=\ds\sqrt{1-\frac{4 m_l^2}{s}}$,
the angle $\vartheta_3$ is the c.m.s. the angle between
$p_1$ and $p_3$.

Then we estimate the cross section as a function of eight helicity amplitudes.
Helicity amplitudes depend on kinematic variables, coupling constants and  seven scalar form factors.
Helicity indices $\mathcal{H}_{\chi_1 \chi_2\chi_3\chi_4}$ denote the signs of the fermion spin projections to corresponding momenta. 
Some basic definitions are
$c^{\pm}=1 \pm \cos\vartheta_3$,
and the scattering angle $\vartheta_3$
is related to the Mandelstam invariants $t,u$:
\bqa
t  &=& m_l^2 - \frac{s}{2} (1 - \beta_l\cos\vartheta_3),
\\
u  &=& m_l^2 - \frac{s}{2} (1 + \beta_l\cos\vartheta_3 ).
\eqa

The presence of the electron masses gives additional terms proportional to
the factor $m_{l}$, which 
can be considered significant in calculations
at low energy.

The set of the corresponding HAs in the $t$ channel
for this case is:
\begin{eqnarray*}
{\cal{H}}_{\mp\mp\mp\mp} &=&
 \frac{1}{t} 
 \Biggl[  (\cpl k-4 \ml^2) \tffgg
+{\chizt}\Bigl(
(\cpl k-4 \ml^2) \tffqq
\nonumber\\
&&
+4 (k \mp \sqrtL) \tffll
 + 2(\cpl k-4 \ml^2 \mp 2 \sqrtL) \tfflq
\nonumber\\
&&
 +  2\cml\ml^2 k \left[\tffqd
 +  (k \mp \sqrtL) \tffld\right]\Bigr)
  \Biggr],
\nonumber   
\end{eqnarray*}

\begin{eqnarray*}
  {\cal{H}}_{\mp--\pm} &=&
  \phantom{-}
      \sin\vartheta_3\frac{\sqrt{s}\ml}{t}
      \Biggl[   \tffgg
     +{\chizt}\Bigl( \tffqq
        + 2 \tfflq 
        - k \tffqd
\nonumber\\
&&
        -(k-\sqrtL )\tffld\Bigr)
            \Biggr],
\nonumber 
\end{eqnarray*}

\begin{eqnarray*}
 {\cal{H}}_{\pm++\mp} &=&  -{\cal{H}}_{\mp--\pm}(\sqrtL \to -\sqrtL),
\end{eqnarray*}

\begin{eqnarray*}
{\cal{H}}_{\mp\mp\pm\pm} &=& 
-\frac{\ml^2}{t}
\Biggl[2 \cpl \tffgg
      + {\chizt}\Bigl( 
 2 \cpl(\tffqq+2\tfflq)
\nonumber\\ &&  
         +8  \tffll-s \cpl (\tffld+\tffqd)\Bigr)\Biggr],
\nonumber   
\end{eqnarray*}

\begin{eqnarray*}
{\cal{H}}_{\mp\pm\mp\pm} &=&
 -\frac{\ml^2}{t}  \Biggl[2 \cpl \tffgg
      + {\chizt}\Bigl( 
 2\cpl(
 \tffqq+2\tfflq)
 \\ \nonumber &&  
 -s \cpl (\tffld+\tffqd)\Bigr)\Biggr],
\nonumber 
\end{eqnarray*}

\begin{eqnarray*}
{\cal{H}}_{-\mp\pm-} &=& -{\cal{H}}_{+\mp\pm+},
\nonumber 
\end{eqnarray*}

\begin{eqnarray*}
{\cal{H}}_{+\mp\pm+} &=&  \phantom{-}
 \sin\vartheta_3 \frac{\sqrt{s}\ml}{t}
 \Biggl[\tffgg
 + {\chizt}\Bigl( \tffqq + 2 \tfflq
 \\ \nonumber &&  
   -2 \ml^2 (\tffld+\tffqd)\Bigr)\Biggr],
\nonumber 
\end{eqnarray*}

\begin{eqnarray*}
{\cal{H}}_{\mp\pm\pm\mp} &=&
      \frac{\cml}{t}
       \Biggl[   k \tffgg
       + {\chizt}\Bigl( 
       k(\tffqq + 2  \tfflq)
       \\ \nonumber &&  
   - 2\ml^2[(k\pm \sqrtL)\tffld + k\tffqd] \Bigr)
       \Biggr].
\end{eqnarray*}

Here  $\chizt$  is the $Z/\gamma$ propagator ratio:
\bqa
\chizt&=&\frac{1}{4{s^2_{\sss{W}}} {c^2_{\sss{W}}}}
\frac{\ds t}{\ds{t - \mzs}}\,.
\label{propagators}
\eqa
Note that {\it tilded} FFs absorb couplings, which leads to a compactification 
of formulas for the amplitude, while explicit expressions will be 
given for {\it untilded} quantities.

The expressions for {\it tilded} FFs are
\begin{eqnarray}
{\tilde F}_{\sss gg} &=& (\alep)^2      {F}_{\sss gg}(s,t,u),\\ \nonumber
{\tilde F}_{\sss ll} &=& \vmalep^2      {F}_{\sss ll}(s,t,u),\\ \nonumber
{\tilde F}_{\sss qq} &=& \alep\vmalep   {F}_{\sss qq}(s,t,u),\\ \nonumber
{\tilde F}_{\sss lq} &=& \vmalep\alep   {F}_{\sss lq}(s,t,u),\\ \nonumber
{\tilde F}_{\sss ql} &=& \vmalep\alep   {F}_{\sss ql}(s,t,u),\\ \nonumber
{\tilde F}_{\sss ld} &=& (\alep)^2      {F}_{\sss ld}(s,t,u),\\ \nonumber
{\tilde F}_{\sss qd} &=&\vmalep\alep    {F}_{\sss qd}(s,t,u).   \nonumber
\end{eqnarray}

 We also use the coupling constants
\bqa
I^{(3)}_l,\; \sigma_l = v_l + a_l,\; \delta_l = v_l - a_l,\;
\stw=\frac{e}{g},\;
\ctw=\frac{\mw}{\mz}\nonumber
\eqa
with $l=e,\mu$.

 In order to get HAs for the Born level, one should set
$F_{\sss gg,ll,lq,ql,qq}=1$ and $F_{\sss ld,qd}=0$.

\subsection{Real photon emission corrections} 

The real corrections consist of soft and hard radiative  contributions.
They are calculated using the bremsstrahlung modules. The soft bremsstrahlung radiation has a Born-like kinematics, 
while the phase space of hard radiation has an extra particle -- photon. 

The \underline{soft bremsstrahlung radiation} has the following form:
\bqa
\sigma^{\rm soft} &=&
       -\qel^2  \frac{2\alpha}{\pi} \sigma^{\rm Born}\bigg[ 
\bigg(1+\frac{1-2 m_l^2/s}{\beta_s}\ln x^2\bigg)\lnlaom   
\nll &&  
     +\frac{1}{\beta_{s}}
     \biggl[-\ln x^2
     + ({1-2 m_l^2/s})
     \bigg(
 \Litwo\left(1-x^2\right)
\nll && 
-\Litwo\left(1-\frac{1}{x^2}\right)\bigg)\biggr]
     -F(t)-F(u)\bigg],
\eqa
were
\bqa
F(I)=    \frac{1-2m_l^2/I}{\beta_{I}}
     \bigg[
     \ln y_{I} \lnlaom 
     +\Litwo\left(\ds 1-\frac{y_{I}x}{z_{I}}\right)
\nll       
     -\Litwo\left(\ds 1-\frac{x}{z_{I}}\right)
     +\Litwo\left(\ds 1-\frac{y_{I}}{z_{I}x}\right)
     -\Litwo\left(\ds 1-\frac{1}{z_{I}x}\right)      
     \bigg], 
     \nonumber \\
\beta_{I} = \sqrt{1-\frac{4m_l^2}{I}}, \quad\quad x = \frac{\sqs}{\ml}\frac{1+\beta_s}{2},
\nll
 y_{I} = 1-\frac{I}{\rlms}\frac{1+\beta_I}{2},\quad
 z_{I} =\frac{\ml}{\sqs} \left(1+y_{I}\right),\quad
\nonumber
\eqa
with $I=t,u$.
 
In presenting the results we used our universal massive module for the
\underline{hard photon bremsstrahlung} for $4f \to 0$ \cite{Bondarenko:2020hhn} by 
appropriate unfolding
it in the right channel.

\section{Numerical results}
\label{Sect_Num}

\subsection{Tree level}

In this Section calculated polarized cross sections at tree level
for the Born and hard photon bremsstrahlung 
are compared with the results of the \CalcHEP{} \cite{Belyaev:2012qa} and \WHIZARD{} 
\cite{Ohl:2006ae, Kilian:2007gr,Kilian:2018onl}
codes.

\begin{table}[!h]
\caption{Table 1.
The tuned triple comparison between \SANC{} (the second column), \WHIZARD{} (the third column), 
and \CalcHEP{} (the fourth column) and \SANC{} results for the hard bremsstrahlung cross section}
\centering	
\begin{tabular}{|l|c|c|c|}
\hline
$P_{e^-},P_{e^-}$ & S         & W         & C \\
\hline
\phantom{$-$}0,\phantom{$-$}0  & 170.12(1) & 170.13(1) & 170.11(2)\\
$-1$,$-1$                      & 284.58(1) & 284.58(1) & 284.55(2)\\
$-1$,\phantom{$-$}1            & 74.00(1)  & 74.00(1)  & 74.00(2)\\
\phantom{$-$}1,$-1$            & 74.01(1)  & 74.02(1)  & 74.00(2)\\
\phantom{$-$}1,\phantom{$-$}1  & 247.90(1) & 247.90(1) & 247.86(2)\\
\hline
\end{tabular}
\label{comphard}
\end{table}

The results are calculated in the $\alpha(0)$ EW scheme
with fixed $100 \%$ polarized initial states 
for $\sqrt{s}=250$~GeV and
angular cuts $|\cos \vartheta_e| \le 0.9$.
For the hard bremsstrahlung cross sections, an additional cut on the
photon energy 
$E_\gamma \ge \omega = 10^{-4} \sqrt{s}/2$
is applied.

The Born results agree in all digits for all codes, and therefore the table is omitted.
The hard bremsstrahlung results are shown in Table~1.
Very good agreement within statistical errors with the above mentioned codes is found.

\subsection{One-loop level}
 
In this Section we show the study of the complete one-loop EW RCs and polarization effects for M{\o}ller scattering in high-energy regions.
Numerical estimates are presented for the total (integrated) cross sections ($\sigma$, pb) and relative corrections ($\delta$, \%) as well as for the 
differential distribution as function of the scattering angle~$\cos\vartheta_{3}$. The channels
$e^-$ ($e^- e^- \to e^- e^- (\gamma$)) and
$\mu^+$ ($\mu^+ \mu^+ \to \mu^+ \mu^+ (\gamma$)) of reaction (\ref{lepton-pm})
are considered below.

\subsubsection{Integrated cross sections}

\texttt{CLIC} would
provide high-luminosity $e^-e^-$ collisions covering a centre-of-mass energy range from 380 GeV to 3 TeV.
They are three main c.m.s. energy stages at ${\sqrt{s}}_{\mathrm{CLIC}}$: 380 GeV, 1.5 TeV and 3 TeV. 

\begin{table}[!h]
\caption{Table 2.
Integrated Born and one-loop cross section in pb
and relative corrections in percent
for $e^-$-channel scattering
for c.m.s. energy $\sqrt{s}_{\mathrm{ILC\&CLIC}}$
and set (\ref{SetPolarization2})  of the initial particle 
polarization degrees in the $\alpha(0)$ EW scheme}
\centering	
\begin{tabular}{|l|l|l|l|}
\hline
$P_{e^-},P_{e^-}$ & 0,~0 & 0.8,~0.8 & $-0.8,-0.8$\\
\hline
\multicolumn{4}{|c|}{$\sqrt{s}$ = 250 GeV} \\
\hline
$\sigma^{\rm Born}$, pb     & 94.661(1)  & 120.152(1) & 136.377(1)\\
$\sigma^{\rm one-loop}$, pb & 103.906(2) & 134.976(2) & 147.224(2)\\
$\delta, \%$                & 9.77(1)    & 12.34(1)   & 7.95(1)\\
\hline
\multicolumn{4}{|c|}{$\sqrt{s}$ = 380 GeV} \\
\hline
$\sigma^{\rm Born}$, pb     & 42.969(1) & 55.739(1) & 65.487(1)\\
$\sigma^{\rm one-loop}$, pb & 47.327(1) & 63.264(1) & 70.345(1)\\
$\delta, \%$                & 10.14(1)  & 13.50(1)  & 7.42(1)\\
\hline
\multicolumn{4}{|c|}{$\sqrt{s}$ = 500 GeV} \\
\hline
$\sigma^{\rm Born}$, pb     & 25.498(1) & 33.430(1) & 39.984(1) \\
$\sigma^{\rm one-loop}$, pb & 28.068(1) & 38.171(2) & 42.627(2) \\
$\delta, \%$                & 10.08(1)  & 14.18(1)  & 6.61(1) \\
\hline
\multicolumn{4}{|c|}{$\sqrt{s}$ =1 TeV} \\
\hline
$\sigma^{\rm Born}$, pb     & 6.657(1) & 8.850(1)  & 10.865(1)\\
$\sigma^{\rm one-loop}$, pb & 7.218(1) & 10.229(1) & 11.104(1)\\
$\delta, \%$                & 8.42(1)  & 15.58(1)  & 2.20(1)\\
\hline
\multicolumn{4}{|c|}{$\sqrt{s}$ = 1.5 TeV} \\
\hline
$\sigma^{\rm Born}$, pb     & 2.992(1) & 3.989(1) & 4.928(1) \\
$\sigma^{\rm one-loop}$, pb & 3.185(1) & 4.635(1) & 4.827(1)\\
$\delta, \%$                & 6.46(1)  & 16.19(1) & $-2.06(1)$ \\
\hline
\multicolumn{4}{|c|}{$\sqrt{s}$ = 3 TeV} \\
\hline
$\sigma^{\rm Born}$, pb     & 0.7536(1) & 1.007(1) & 1.249(1)\\
$\sigma^{\rm one-loop}$, pb & 0.7665(1) & 1.177(1) & 1.103(1)\\
$\delta, \%$                & 1.71(1)   & 16.94(1) & $-11.70(1)$\\
\hline
\end{tabular}
\label{em_table}
\end{table}

The \texttt{ILC}  offers many opportunities for measurements 
with collider energies from 90 GeV to 1 TeV. 
Three main c.m.s. energy stages 
can be distinguished --
${\sqrt{s}}_{\mathrm{ILC}}$:~250 GeV, 500 GeV, and 1 TeV, with electron polarization of
$P_{e^-} = \pm 0.8$. 

Table 2 presents the integrated Born and one-loop cross section in pb
and relative corrections in percent
for the $e^-$-channel
for c.m.s. energy $\sqrt{s}_{\mathrm{ILC\&CLIC}}$
and set (\ref{SetPolarization2})  of the initial particle
polarization degrees in the $\alpha(0)$ EW scheme.

Under the $\mu$TRISTAN{} experimental conditions, the energy is assumed to be $\sqrt{s}_{\mu\mathrm{TRISTAN}}$:~0.6, 1, 2 TeV and the polarization of both beams will reach
$P_{\mu^+} = \pm 0.8$ for the $\mu^+$-channel.
Table 3 presents the same observables as in Table 2 
in the conditions of the $\mu$TRISTAN experiment.

\begin{table}[!h]
\caption{Table 3.
Integrated Born and one-loop cross section in pb
and relative corrections in percent
for the $\mu^+$-channel
for c.m.s. energy $\sqrt{s}_{\mu\mathrm{TRISTAN}}$
and set (\ref{SetPolarization2})  of the initial particle 
polarization degrees in the $\alpha(0)$ EW scheme}
\centering
\begin{tabular}{|l|l|l|l|}
\hline
$P_{\mu^+},P_{\mu^+}$ & 0,~0 & 0.8,~0.8 & $-0.8,-0.8$\\
\hline
\multicolumn{4}{|c|}{$\sqrt{s}$ = 600 GeV} \\
\hline
$\sigma^{\rm Born}$, pb     & 17.974(1) & 23.690(1) & 28.601(1)\\
$\sigma^{\rm one-loop}$, pb & 19.715(1) & 27.064(1) & 30.160(1)\\
$\delta, \%$                & 9.69(1)   & 14.24(1)  & 5.45(1)\\
\hline
\multicolumn{4}{|c|}{$\sqrt{s}$ = 1 TeV} \\
\hline
$\sigma^{\rm Born}$, pb     & 6.6572(1) & 8.8497(1)  & 10.8648(1) \\
$\sigma^{\rm one-loop}$, pb & 7.2019(1) & 10.1930(1) & 11.0589(2) \\
$\delta, \%$                & 8.18(1)   & 15.18(1)   & 1.79(1)\\
\hline
\multicolumn{4}{|c|}{$\sqrt{s}$ = 2 TeV} \\
\hline
$\sigma^{\rm Born}$, pb     & 1.6903(1) & 2.2559(1) & 2.7935(1)\\
$\sigma^{\rm one-loop}$, pb & 1.7646(1) & 2.6195(1) & 2.6210(1) \\
$\delta, \%$                & 4.40(1)   & 16.12(1)  & $-6.17(1)$\\
\hline
\end{tabular}
\label{mp_table}
\end{table}

As it seen in Tables 2 and 3, the use of the polarized beams
significantly increases the cross section. At the same time the
RCs increase at $P_{e^-} = (0.8,0.8)$ and reduce at $P_{e^-} = (-0.8,-0.8)$  
comparing to the unpolarized case in the region of c.m.s. energies 
$\sqrt{s} = 250-1000$ GeV. At higher c.m.s. energies 
the polarization $P_{e,\mu} = (\pm 0.8,\pm 0.8)$ increases the
cross section as well, but the absolute value of the relative 
correction became larger than for the unpolarized one.

\subsubsection{Differential cross sections}

At Figures 1 and 2 the differential distributions for the LO and EW NLO cross sections (in pb) as well as the relative corrections (in \%) are shown for the $e^-$- and $\mu^+$-channels for the c.m.s. energies $\sqrt{s}=250, 1000$~GeV as a function of $\cos \theta_3$.

The differential distributions over $\cos \theta_3$ are symmetric for all c.m.s. energies and the maximum of the relative corrections is at the zero angle while the minimum is close to the $|\cos \theta_3| = 1$. This is due to dominance of the Born contribution in the $|\cos \theta_3| \approx 1$ region due to a photon propagator $1/t (1/u)$.

It should be noted that while the integrated relative corrections for the c.m.s. energy $\sqrt{s} = 1000$ GeV~for the $e^-$- and $\mu^+$-channels differ by $0.4\%$ (see Tables 2,3) the  differential difference is larger, being about 5-6 \% at $\cos\theta_3 = 0$.

\begin{figure}[ht]
\centering
\includegraphics[width=0.7\textwidth]{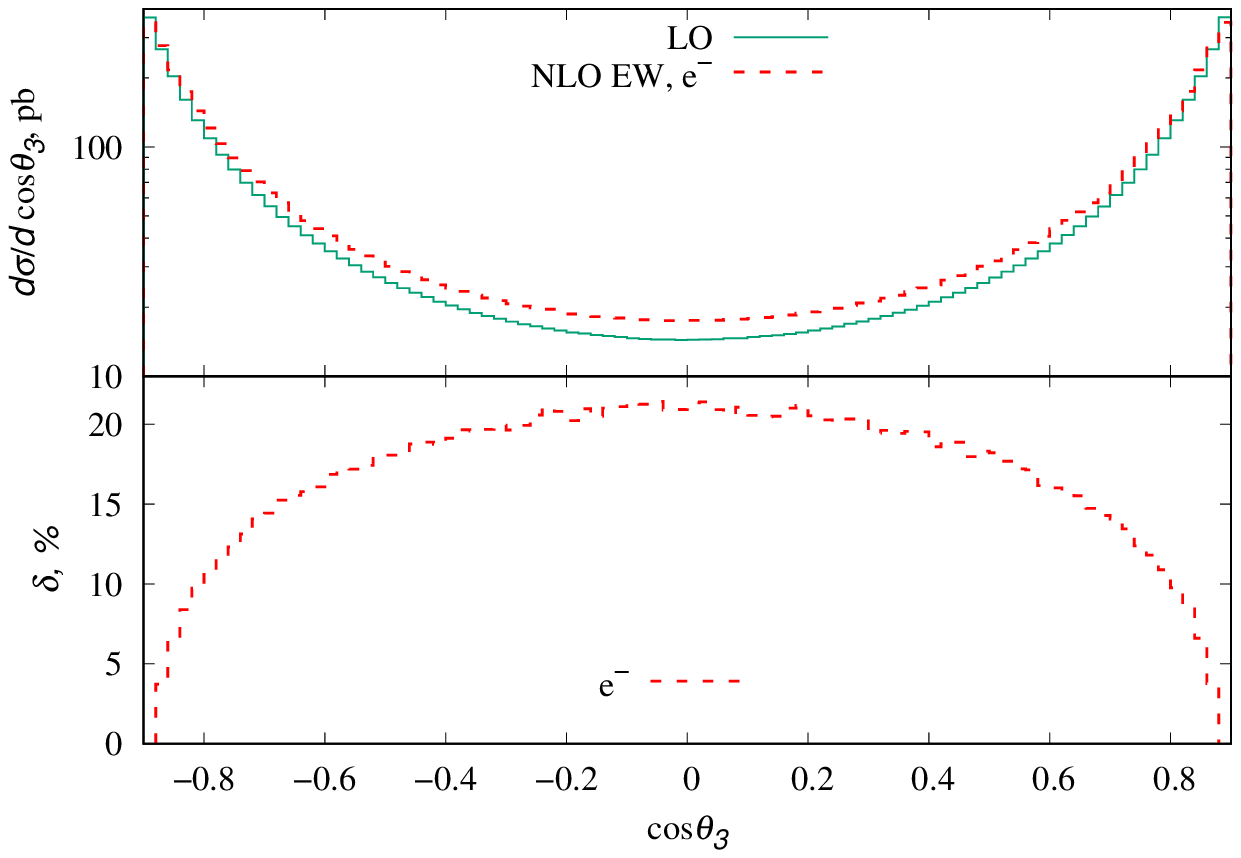}
\caption{Fig. 1. The LO and NLO EW unpolarized cross sections (upper panel) and relative corrections (lower panel)
of the $e^-$-channel
for the c.m.s. energy $\sqrt{s}=250$~GeV
as a function of $\cos \theta_3$}
\label{fig:emem-cos3-250}
\end{figure}

\begin{figure}[ht]
\centering
\includegraphics[width=0.7\textwidth]{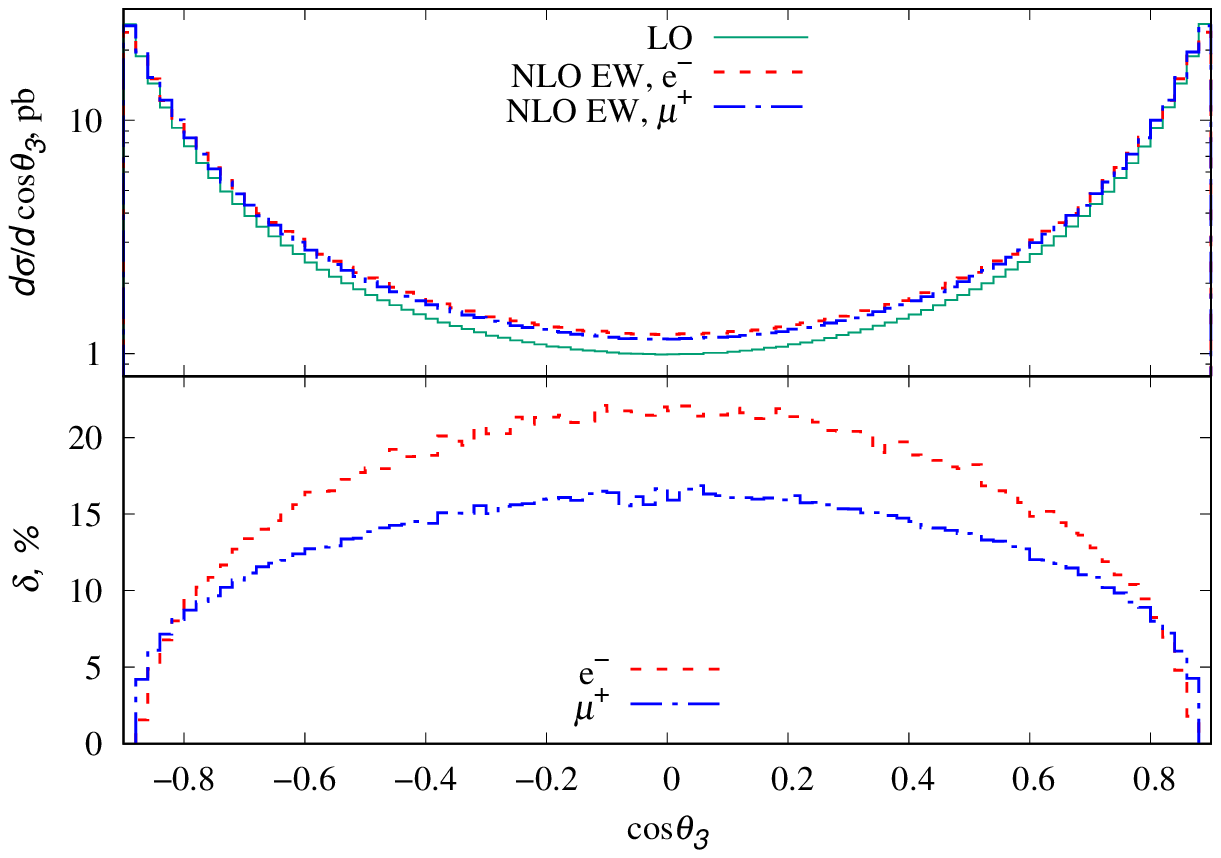}
\caption{Fig. 2. The LO and NLO EW unpolarized cross sections (upper panel) and relative corrections (lower panel)
of the $e^-$- and $\mu^+$-channels
for the c.m.s. energy $\sqrt{s}=1000$~GeV
as a function of $\cos \theta_3$}
\label{fig:emem-cos3-1000}
\end{figure}

\section{Summary}\label{Sect_C}

We computed the NLO contributions RCs
due to QED and purely weak corrections and implement them into a fully differential
Monte Carlo event generator \ReneSANCe{}~
and \MCSANC{} integrator. 

We presented explicit expressions for HAs to evaluate
virtual and soft parts for M{\o}ller scattering.
We  used our previous module for HAs of the hard photon 
bremsstrahlung~\cite{Bondarenko:2020hhn}.
We showed  the results of interest for 
unpolarized~\texttt{FCCee} and 
polarized  \texttt{ILC}, \texttt{CLIC}, $\mu$\texttt{TRISTAN}
experiments.

Our results show that
it is necessary to include more then one-loop corrections to ensure the required level of the theoretical support. 

The established \SANC~ framework allows us to investigate
the one-loop and higher-order corrections
for any polarization, estimate the contribution of the selected  helicity state, and  take into account mass effects.

\section{Funding}
This research was
supported by the Russian Foundation for Basic Research, project
N 20-02-00441.
 

\begingroup\begin{thebibliography}{10}

\bibitem{homepagesILC}
ILC homepages --- {\em https://www.linearcollider.org/ILC}.

\bibitem{Irles:2019xny}
A.~Irles, R.~Poschl, F.~Richard, and H.~Yamamoto, ``{Complementarity between
  ILC250 and ILC-GigaZ}'', in {\em {Linear Collider Community Meeting Lausanne,
  Switzerland, April 8-9, 2019}}, 2019,
\href{http://www.arXiv.org/abs/1905.00220}{{\tt 1905.00220}}.

\bibitem{Moortgat-Picka:2015yla}
A.~Arbey {\em et al.}, {\em Eur. Phys. J.} {\bf C75} (2015), no.~8 371,
\href{http://www.arXiv.org/abs/1504.01726}{{\tt 1504.01726}}.

\bibitem{Baer:2013cma}
H.~Baer, T.~Barklow, K.~Fujii, Y.~Gao, A.~Hoang, S.~Kanemura, J.~List, H.~E.
  Logan, A.~Nomerotski, M.~Perelstein, {\em et al.},
\href{http://www.arXiv.org/abs/1306.6352}{{\tt 1306.6352}}.

\bibitem{Accomando:1997wt}
{ECFA/DESY LC Physics Working Group} Collaboration, E.~Accomando {\em et al.},
  {\em Phys. Rept.} {\bf 299} (1998) 1--78,
\href{http://www.arXiv.org/abs/hep-ph/9705442}{{\tt hep-ph/9705442}}.

\bibitem{Battaglia:2004mw}
{CLIC Physics Working Group} Collaboration, E.~Accomando {\em et al.},
  ``{Physics at the CLIC multi-TeV linear collider}'', in {\em {Proceedings,
  11th International Conference on Hadron spectroscopy (Hadron 2005): Rio de
  Janeiro, Brazil, August 21-26, 2005}}, 2004,
\href{http://www.arXiv.org/abs/hep-ph/0412251}{{\tt hep-ph/0412251}}.

\bibitem{homepagesFCCee}
FCC-ee homepages --- {\em http://tlep.web.cern.ch}.

\bibitem{Abada:2019ono}
{FCC} Collaboration, A.~Abada {\em et al.}, {\em Eur. Phys. J. ST} {\bf 228}
  (2019), no.~5
1109--1382.

\bibitem{Abada:2019lih}
{FCC} Collaboration, A.~Abada {\em et al.}, {\em Eur. Phys. J.} {\bf C79}
  (2019), no.~6
474.

\bibitem{Blondel:2019ykp}
A.~Blondel and P.~Janot,
\href{http://www.arXiv.org/abs/1912.11871}{{\tt 1912.11871}}.

\bibitem{Blondel:2018mad}
A.~Blondel {\em et al.}, ``{Standard model theory for the FCC-ee Tera-Z
  stage}'', in {\em {Mini Workshop on Precision EW and QCD Calculations for the
  FCC Studies : Methods and Techniques CERN, Geneva, Switzerland, January
  12-13, 2018}}, vol.~3, CERN, CERN, Geneva, 2019,
\href{http://www.arXiv.org/abs/1809.01830}{{\tt 1809.01830}}.

\bibitem{homepagesCLIC}
CLIC homepages --- {\em http://clic-study.web.cern.ch}.

\bibitem{CLIC:2016zwp}
{CLIC, CLICdp} Collaboration, M.~J. Boland {\em et al.},
  \href{http://www.arXiv.org/abs/1608.07537}{{\tt 1608.07537}}.

\bibitem{Charles:2018vfv}
{CLICdp, CLIC} Collaboration, T.~K. Charles {\em et al.}, {\em CERN Yellow Rep.
  Monogr.} {\bf 1802} (2018) 1--98,
\href{http://www.arXiv.org/abs/1812.06018}{{\tt 1812.06018}}.

\bibitem{homepagesCEPC}
CEPC homepages --- {\em http://cepc.ihep.ac.cn}.

\bibitem{Fujii:2020pxe}
K.~Fujii {\em et al.}, \href{http://www.arXiv.org/abs/2007.03650}{{\tt
  2007.03650}}.

\bibitem{Hamada:2022mua}
Y.~Hamada, R.~Kitano, R.~Matsudo, H.~Takaura, and M.~Yoshida,
  \href{http://www.arXiv.org/abs/2201.06664}{{\tt 2201.06664}}.

\bibitem{Moller1932}
C.~Møller, {\em Annalen der Physik} {\bf 14} (1932).

\bibitem{Jadach:1994im}
S.~Jadach and B.~F.~L. Ward, {\em Phys. Rev. D} {\bf 54} (1996) 743--749.

\bibitem{Shumeiko:1999zd}
N.~M. Shumeiko and J.~G. Suarez, {\em J. Phys. G} {\bf 26} (2000) 113--127,
  \href{http://www.arXiv.org/abs/hep-ph/9912228}{{\tt hep-ph/9912228}}.

\bibitem{Montero:1998ve}
J.~C. Montero, V.~Pleitez, and M.~C. Rodriguez, {\em Phys. Rev. D} {\bf 58}
  (1998) 094026, \href{http://www.arXiv.org/abs/hep-ph/9802313}{{\tt
  hep-ph/9802313}}.

\bibitem{Denner:1998um}
A.~Denner and S.~Pozzorini, {\em Eur. Phys. J. C} {\bf 7} (1999) 185--195,
  \href{http://www.arXiv.org/abs/hep-ph/9807446}{{\tt hep-ph/9807446}}.

\bibitem{Czarnecki:2000ic}
A.~Czarnecki and W.~J. Marciano, {\em Int. J. Mod. Phys. A} {\bf 15} (2000)
  2365--2376, \href{http://www.arXiv.org/abs/hep-ph/0003049}{{\tt
  hep-ph/0003049}}.

\bibitem{Alexander:2000bu}
G.~Alexander and I.~Cohen, {\em Nucl. Instrum. Meth. A} {\bf 486} (2002)
  552--567, \href{http://www.arXiv.org/abs/hep-ex/0006007}{{\tt
  hep-ex/0006007}}.

\bibitem{Petriello:2002wk}
F.~J. Petriello, {\em Phys. Rev. D} {\bf 67} (2003) 033006,
  \href{http://www.arXiv.org/abs/hep-ph/0210259}{{\tt hep-ph/0210259}}.

\bibitem{Ilyichev:2005rx}
A.~Ilyichev and V.~Zykunov, {\em Phys. Rev. D} {\bf 72} (2005) 033018,
  \href{http://www.arXiv.org/abs/hep-ph/0504191}{{\tt hep-ph/0504191}}.

\bibitem{MOLLER:2014iki}
{MOLLER} Collaboration, J.~Benesch {\em et al.},
  \href{http://www.arXiv.org/abs/1411.4088}{{\tt 1411.4088}}.

\bibitem{Aleksejevs:2010ub}
A.~Aleksejevs, S.~Barkanova, A.~Ilyichev, and V.~Zykunov, {\em Phys. Rev. D}
  {\bf 82} (2010) 093013, \href{http://www.arXiv.org/abs/1008.3355}{{\tt
  1008.3355}}.

\bibitem{Ahmadov:2012se}
A.~I. Ahmadov, Y.~M. Bystritskiy, E.~A. Kuraev, A.~N. Ilyichev, and V.~A.
  Zykunov, {\em Eur. Phys. J. C} {\bf 72} (2012) 1977,
  \href{http://www.arXiv.org/abs/1201.0460}{{\tt 1201.0460}}.

\bibitem{Aleksejevs:2015zya}
A.~G. Aleksejevs, S.~G. Barkanova, Y.~M. Bystritskiy, E.~A. Kuraev, and V.~A.
  Zykunov, {\em Phys. Part. Nucl. Lett.} {\bf 13} (2016), no.~3 310--317,
  \href{http://www.arXiv.org/abs/1508.07853}{{\tt 1508.07853}}.

\bibitem{Akushevich:2015toa}
I.~Akushevich, H.~Gao, A.~Ilyichev, and M.~Meziane, {\em Eur. Phys. J. A} {\bf
  51} (2015), no.~1 1.

\bibitem{Krauss:2001iv}
F.~Krauss, R.~Kuhn, and G.~Soff, {\em JHEP} {\bf 02} (2002) 044,
  \href{http://www.arXiv.org/abs/hep-ph/0109036}{{\tt hep-ph/0109036}}.

\bibitem{Belyaev:2012qa}
A.~Belyaev, N.~D. Christensen, and A.~Pukhov, {\em Comput. Phys. Commun.} {\bf
  184} (2013) 1729--1769,
\href{http://www.arXiv.org/abs/1207.6082}{{\tt 1207.6082}}.

\bibitem{Yuasa:1999rg}
F.~Yuasa {\em et al.}, {\em Prog. Theor. Phys. Suppl.} {\bf 138} (2000) 18--23,
  \href{http://www.arXiv.org/abs/hep-ph/0007053}{{\tt hep-ph/0007053}}.

\bibitem{Belanger:2003sd}
G.~Belanger, F.~Boudjema, J.~Fujimoto, T.~Ishikawa, T.~Kaneko, K.~Kato, and
  Y.~Shimizu, {\em Phys. Rept.} {\bf 430} (2006) 117--209,
  \href{http://www.arXiv.org/abs/hep-ph/0308080}{{\tt hep-ph/0308080}}.

\bibitem{Kilian:2007gr}
W.~Kilian, T.~Ohl, and J.~Reuter, {\em Eur. Phys. J.} {\bf C71} (2011) 1742,
\href{http://www.arXiv.org/abs/0708.4233}{{\tt 0708.4233}}.

\bibitem{Afanasev:2006xs}
A.~Afanasev, E.~Chudakov, A.~Ilyichev, and V.~Zykunov, {\em Comput. Phys.
  Commun.} {\bf 176} (2007) 218--231,
  \href{http://www.arXiv.org/abs/hep-ph/0603027}{{\tt hep-ph/0603027}}.

\bibitem{Bardin:2017mdd}
D.~Bardin, Y.~Dydyshka, L.~Kalinovskaya, L.~Rumyantsev, A.~Arbuzov, R.~Sadykov,
  and S.~Bondarenko, {\em Phys. Rev. D} {\bf 98} (2018), no.~1 013001,
  \href{http://www.arXiv.org/abs/1801.00125}{{\tt 1801.00125}}.

\bibitem{Bondarenko:2018sgg}
S.~Bondarenko, Y.~Dydyshka, L.~Kalinovskaya, L.~Rumyantsev, R.~Sadykov, and
  V.~Yermolchyk, {\em Phys. Rev. D} {\bf 100} (2019), no.~7 073002,
  \href{http://www.arXiv.org/abs/1812.10965}{{\tt 1812.10965}}.

\bibitem{Bondarenko:2020hhn}
S.~Bondarenko, Y.~Dydyshka, L.~Kalinovskaya, R.~Sadykov, and V.~Yermolchyk,
  {\em Phys. Rev. D} {\bf 102} (2020), no.~3 033004,
  \href{http://www.arXiv.org/abs/2005.04748}{{\tt 2005.04748}}.

\bibitem{Bondarenko:2021eni}
S.~Bondarenko, Y.~Dydyshka, L.~Kalinovskaya, L.~Rumyantsev, R.~Sadykov, and
  V.~Yermolchyk, \href{http://www.arXiv.org/abs/2111.11490}{{\tt 2111.11490}}.

\bibitem{Bondarenko:2022ddm}
S.~Bondarenko, L.~Kalinovskaya, and A.~Sapronov,
  \href{http://www.arXiv.org/abs/2201.04350}{{\tt 2201.04350}}.

\bibitem{Sadykov:2020any}
R.~Sadykov and V.~Yermolchyk, {\em Comput. Phys. Commun.} {\bf 256} (2020)
  107445, \href{http://www.arXiv.org/abs/2001.10755}{{\tt 2001.10755}}.

\bibitem{Passarino:1978jh}
G.~Passarino and M.~J.~G. Veltman, {\em Nucl. Phys. B} {\bf 160} (1979)
  151--207.

\bibitem{Andonov:2002xc}
A.~Andonov, D.~Bardin, S.~Bondarenko, P.~Christova, L.~Kalinovskaya, and
  G.~Nanava, {\em Phys. Part. Nucl.} {\bf 34} (2003) 577--618, [Fiz. Elem.
  Chast. Atom. Yadra 34,1125(2003)],
\href{http://www.arXiv.org/abs/hep-ph/0207156}{{\tt hep-ph/0207156}}.

\bibitem{Ohl:2006ae}
T.~Ohl, ``{WHiZard and O'Mega}'', in {\em {Proceedings, LoopFest V: Radiative
  Corrections for the International Linear Collider: Multi-loops and
  Multi-legs: SLAC, Menlo Park, California, June 19-21, 2006}},
2006.

\bibitem{Kilian:2018onl}
W.~Kilian, S.~Brass, T.~Ohl, J.~Reuter, V.~Rothe, P.~Stienemeier, and M.~Utsch,
  ``{New Developments in WHIZARD Version 2.6}'', in {\em {International
  Workshop on Future Linear Collider (LCWS2017) Strasbourg, France, October
  23-27, 2017}}, 2018,
\href{http://www.arXiv.org/abs/1801.08034}{{\tt 1801.08034}}.

\end{thebibliography}\endgroup

\providecommand{\href}[2]{#2}

\end{document}